\documentclass[aps,prl,twocolumn,superscriptaddress,longbibliography]{revtex4-1}
\usepackage{pdfpages}
\usepackage{times}
\usepackage{float}
\usepackage{multirow}
\usepackage{amsmath}
\usepackage{amsthm}
\usepackage{amssymb}
\usepackage{amsbsy}
\usepackage[english]{babel}
\usepackage[T1]{fontenc}
\usepackage[utf8]{inputenc}
\usepackage{graphicx}
\usepackage[colorlinks,bookmarks=false,citecolor=blue,linkcolor=red,urlcolor=blue]{hyperref}
\usepackage{xcolor}
\usepackage{pstricks}
\usepackage{tabularx,hhline}
\usepackage[caption=false]{subfig}
\graphicspath{{./figs/}} 

\makeatletter
\AtBeginDocument{\let\LS@rot\@undefined}
\makeatother

\newcommand{\CCO}{Ca$_{10}$Cr$_7$O$_{28}$}

\begin{document}

\title{How many spin liquids are there in Ca$_{10}$Cr$_7$O$_{28}$?}

\author{Rico Pohle}
\email{rico.pohle@oist.jp}
\affiliation{Okinawa Institute of Science and Technology Graduate University, Onna-son, 
	Okinawa 904-0412, Japan}

\author{Han Yan}
\email{han.yan@oist.jp}
\affiliation{Okinawa Institute of Science and Technology Graduate University, Onna-son, 
	Okinawa 904-0412, Japan}

\author{Nic Shannon}
\email{nic.shannon@oist.jp}
\affiliation{Okinawa Institute of Science and Technology Graduate University, Onna-son, 
	Okinawa 904-0412, Japan}

\date{\today}

\begin{abstract}


The search for novel phases of matter is a central theme of modern
physics, with some of the most intriguing examples provided by the 
spin liquids found in magnets with competing, or ``frustrated'' 
interactions.
\CCO, a novel spin--$1/2$ magnet with a bilayer breathing--kagome 
lattice, has properties which differ from from any known spin liquid.
However, understanding \CCO\ presents a significant 
challenge, because of its complex frustration.
Here we use large--scale molecular--dynamics simulation to 
explore the origin of spin--liquid behaviour in \CCO.
We uncover qualitatively different behaviour on different timescales, 
and argue that ground state of \CCO\ is born out of 
a slowly--fluctuating ``spiral spin liquid'', while faster fluctuations 
echo the U(1) spin liquid found in the kagome antiferromagnet.
These results provide a concrete scenario for spin--liquid behaviour in
\CCO, and highlight the possibility of spin liquids existing on 
multiple timescales.   

\end{abstract}


\maketitle


The search for novel quantum phases and excitations is one of the 
defining themes of modern physics, reaching from the large--scale 
structure of the universe, 
to many--body physics on the scale of few Angstrom.
%
Quantum spin systems, and in particular, magnets with competing 
or ``frustrated'' interactions, occupy a special place in this endeavour, 
and have consistently proved to be one of the most fruitful places to 
look for new phases of matter.
While conventional magnets order at low temperatures, 
and exhibit excitations with the character of spin waves,   
frustrated magnets display a far richer behaviour.
In particular, they can support a {\it quantum spin liquid} (QSL), a 
massively--entangled phase of matter in which evades all conventional forms of 
magnetic order, and instead supports entirely new forms of excitation, 
often with topological character \cite{balents10-Nature464}.
The discussion of quantum spin liquids has a long history 
\cite{anderson73-MatResBull8}, but until recently experimental 
realisations remained scarce \cite{lee08-Science321}.
Happily, however, the past few years have seen 
an explosion in the number systems under study, with examples including 
quasi--2D organics \cite{lee08-Science321,shimizu03-PRL91}, 
thin films of $^3$He \cite{masutomi04-PRL92}, 
spin--1/2 magnets with a Kagome lattice \cite{han12-Nature492}, 
``Kitaev'' magnets with strongly anisotropic exchange 
\cite{kitaev06-AnnPhys321,jackeli09-PRL102,banerjee16-NatMater15,baek17-PRL119},  
and quantum analogues of spin ice 
\cite{hermele04-PRB69,banerjee08-PRL100,benton12-PRB86,sibille-arXiv}.   

\begin{figure}[h]
	\captionsetup[subfigure]{labelformat=empty, farskip=2pt,captionskip=1pt}
	\centering	
	\subfloat[(a) $E =$ 0.9 meV \label{fig:Exp_090meV}]{
		\includegraphics[width=0.5\columnwidth]{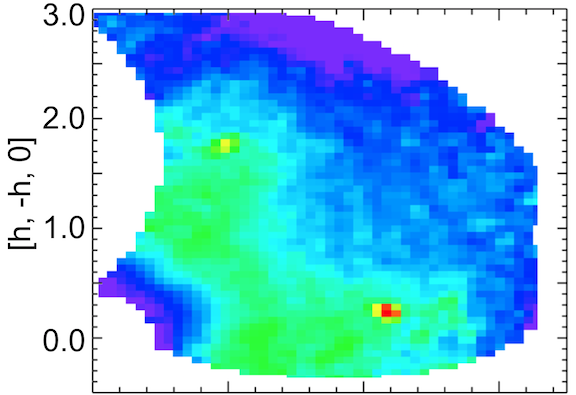}
		}
	\subfloat[ (b) $E =$ 1.1 meV \label{fig:MD_110meV}]{
  		\includegraphics[width=0.45\columnwidth]{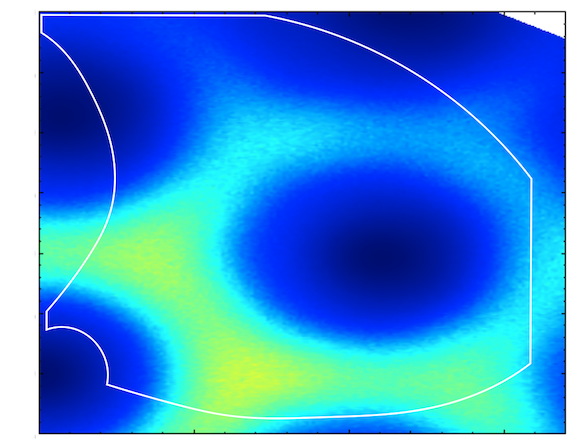}
	}
	\\[1ex]
	
	\subfloat[(c) $E =$ 0.65 meV \label{fig:Exp_065meV}]{
  		\includegraphics[width=0.5\columnwidth]{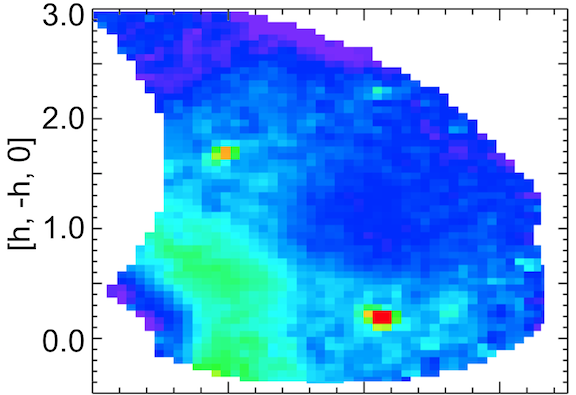}
	}
	\subfloat[(d) $E =$ 0.52 meV  \label{fig:MD_52meV}]{ 
  		\includegraphics[width=0.45\columnwidth]{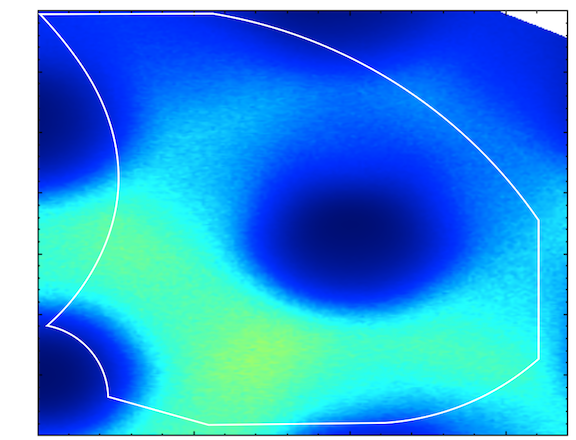}
	}
	\\[1ex]

	\subfloat[(e) $E =$ 0.25 meV  \label{fig:Exp_025meV}]{
  		\includegraphics[width=0.5\columnwidth]{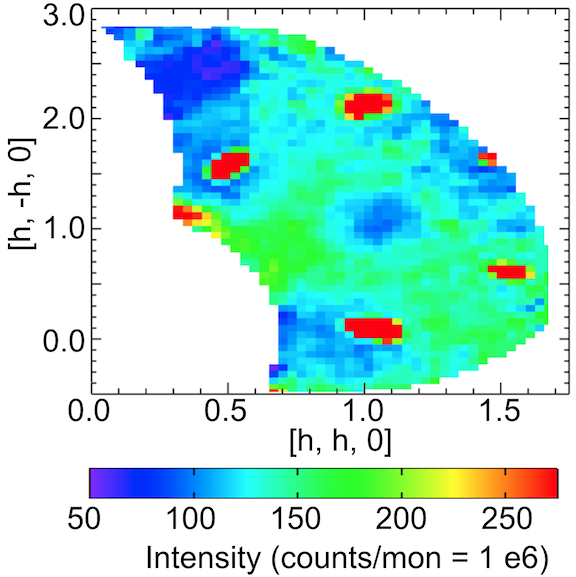}
	}
	\subfloat[(f) $E =$ 0.23 meV 	\label{fig:MD_023meV}]{ 
  		\includegraphics[width=0.45\columnwidth]{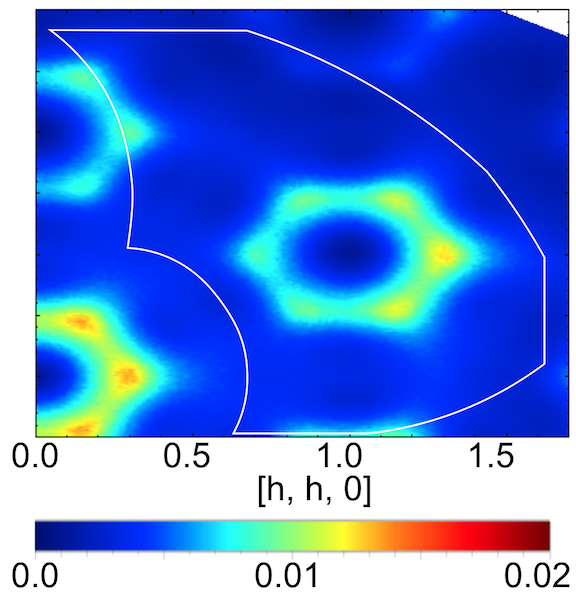}
	}
	\caption{
	{\bf Fluctuations in the spin--liquid phase of \CCO, showing 
	${\bf q}$--dependent structure on different timescales.}
	({\bf a})~Cut through inelastic neutron 
	scattering (INS) data for \CCO\ at high energy, 
	showing ``bow--tie'' structure centered on $(0.5, 0.5, 0)$.    
	({\bf b})~Equivalent results 
	from molecular dynamics (MD) simulations of 
	${\cal H}_{\sf BBK}$ [Eq.~(\ref{eq:H.BBK})].	
	({\bf c})~INS data for \CCO\ at intermediate energy, 
	also  showing ``bow--tie'' structure.
        ({\bf d})~Equivalent results from MD simulation.   
	({\bf e})~INS data for \CCO, suggesting a ``ring''
	of scattering at low energy.      
        ({\bf f})~Equivalent results from MD simulation.   
        Experimental data is reproduced from 
        \cite{balz16-NatPhys12}; 
        details of simulations are given in the text.   
        INS experiments were carried out at $T = 90\ \text{mK}$; 
        MD simulations at $T = 222\ \text{mK}$.
 	}
	\label{fig:EnergySlicesB_0T}
\end{figure}

An exciting new arrival on this scene is 
the quasi--2D magnet \CCO, a system which appears to have qualitatively different 
properties from any previously--studied spin liquid.
First studied for its unusual chemistry \cite{arcon98-JAmCeramSoc81,gyepesova13-ActaCrysC69}, 
recent experiments 
identified \CCO\ as a system where spin--$1/2$ Cr$^{5+}$ ions occupy sites of a bilayer 
breathing Kagome (BBK) lattice with extremely complex frustration 
\cite{balz16-NatPhys12,balz17-PRB95,balz17-JPCM29}.    
%
Heat--capacity, neutron--scattering and $\mu$SR experiments on \CCO, 
find no traces of magnetic order down to a temperature of $19\ \text{mK}$, two orders 
of magnitude lower than the scale of interactions \cite{balz16-NatPhys12}.
Meanwhile, inelastic neutron scattering \cite{balz16-NatPhys12,balz17-PRB95} 
reveals ``bow--tie'' like structure at intermediate to high energies 
[Fig.~\ref{fig:Exp_090meV},\ref{fig:Exp_065meV}], and qualitatively different scattering, 
more reminiscent of a ring, at lower energies [Fig.~\ref{fig:Exp_025meV}].  
%
Parallel pseudo--fermion functional renormalisation group calculations suggest 
that the ground state of \CCO\ should be a quantum spin liquid, characterised 
by a ``ring'' in the static structure factor $S({\bf q}, \omega=0)$ \cite{balz16-NatPhys12}.   
None the less, 
the origin of this spin liquid, and the attendant ring in scattering, remain obscure 
\cite{balz16-NatPhys12,balz17-PRB95,balz17-JPCM29}.  


In this Article we address the nature and origin of spin liquid behaviour in \CCO, 
starting from the microscopic model introduced by 
Balz {\it et al.}~\cite{balz16-NatPhys12, balz17-PRB95,balz17-JPCM29}, 
and using large--scale molecular dynamics 
simulation to explore its semi--classical spin dynamics.  
We find that the spins continue to fluctuate down to very low temperatures, 
and that fluctuations on different timescales encode signatures usually 
associated with two distinct types of spin liquid; a slowly--fluctuating 
``spiral spin liquid'' characterised by a degenerate ``ring'' of spin 
configurations; and, on shorter timescales, ``bow--tie'' structure reminiscent 
of the pinch points observed in the Kagome-lattice antiferromagnet.  
Applying a magnetic field opens a gap to transverse spin 
excitations; we argue that the quantum spin--liquid ground state 
in \CCO\ is born out of the slow fluctuations associated with the 
degenerate ``ring'' at a field of $1\ \text{T}$, where this gap closes.  
We also offer a simple explanation of this degenerate ring, through 
a mapping on to an effective spin--$3/2$ Heisenberg model on a 
honeycomb lattice.   
%
These results provide a concrete scenario for spin-liquid behaviour in \CCO, 
and highlight the possibility of different spin liquids supporting qualitatively 
different properties on different timescales.



\begin{figure*}[t]
	\centering	
	\hspace{-1.5cm}\begin{minipage}[t]{0.6\textwidth}
	\centering
	\subfloat[BBK lattice\label{fig:BBK.model}]{
  		\includegraphics[height=4.8cm]{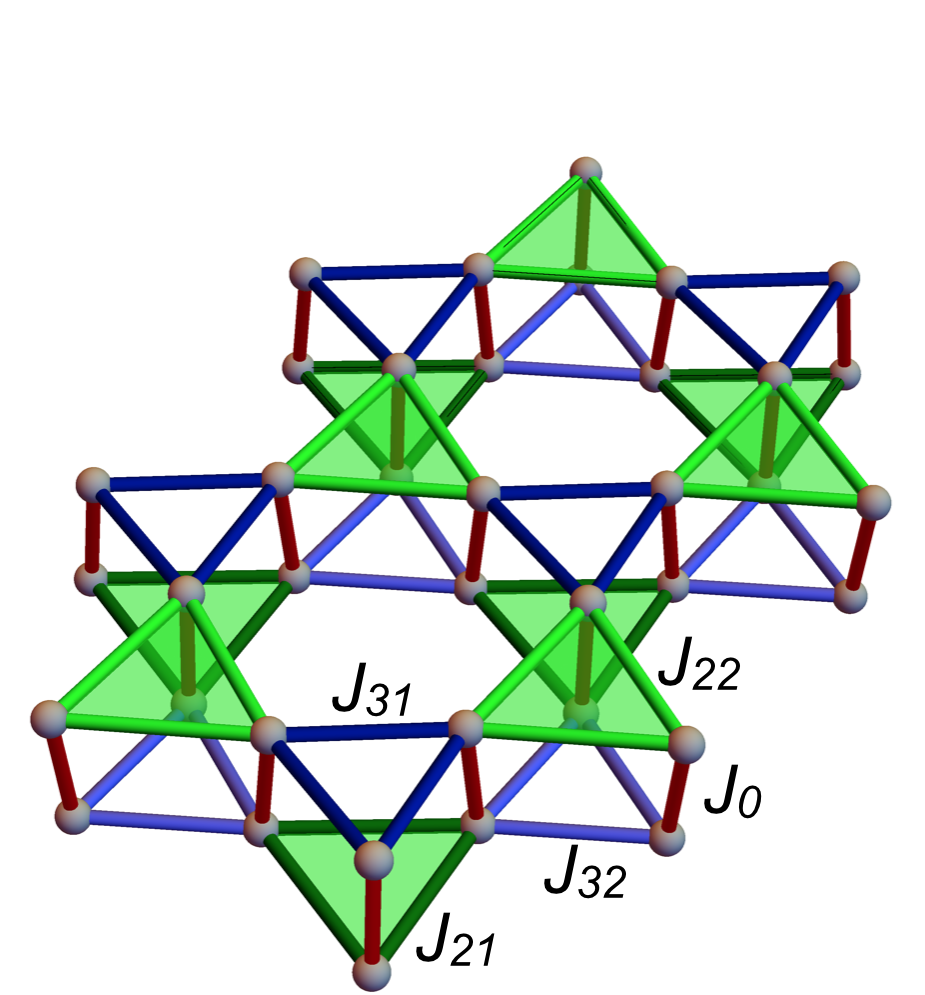}
	}
	\subfloat[Honeycomb lattice\label{fig:HCM.model}]{
  		\includegraphics[height=4.8cm]{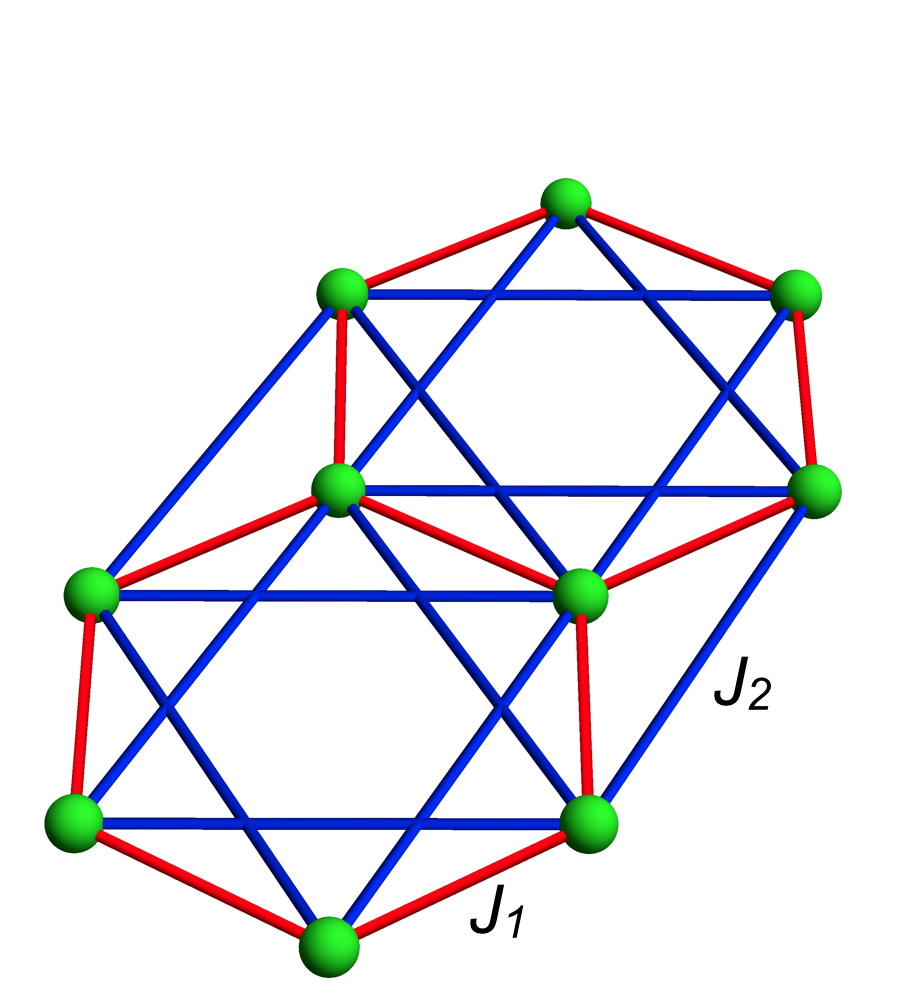}
	}
	\end{minipage}
	\begin{minipage}[t]{0.35\textwidth}
	\centering
	\subfloat[BBK model, $T > 0$\label{fig:phase.diagram.BBK}]{
  		\includegraphics[width=7cm]{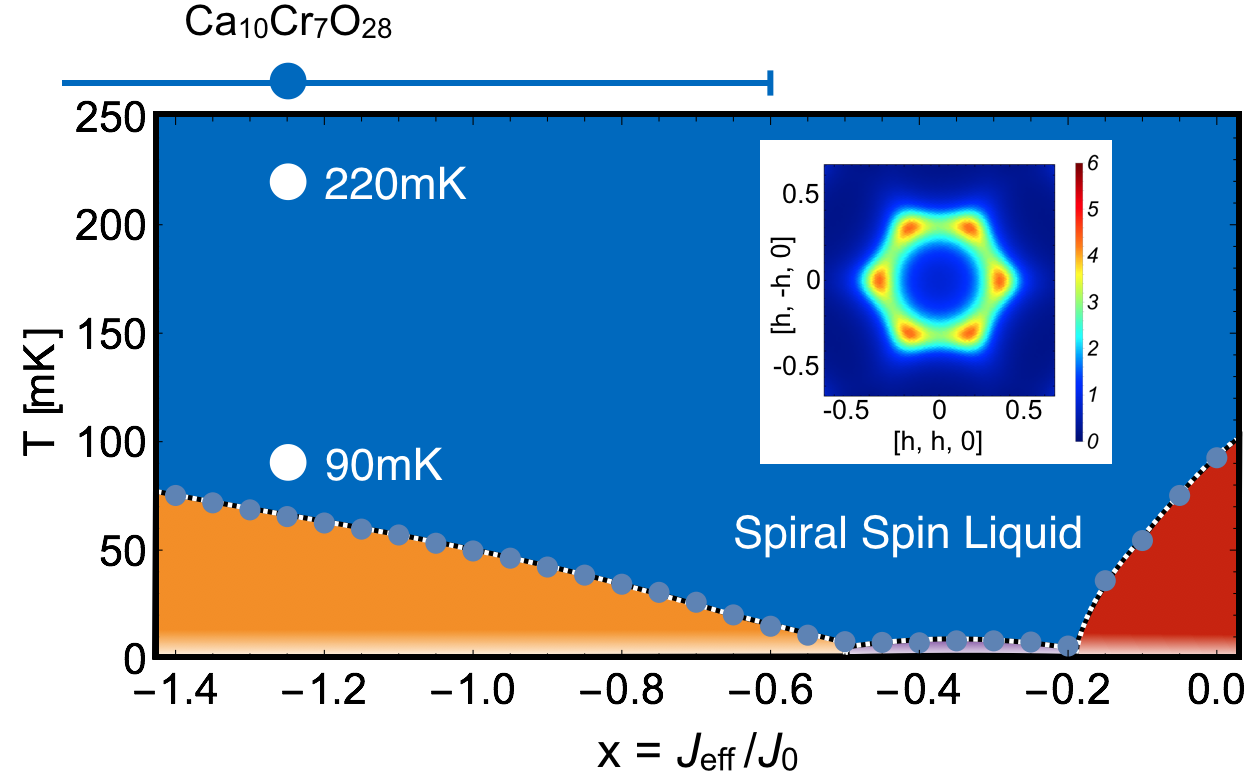}
	}\\
	\subfloat[Honeycomb model, $T = 0$\label{fig:phase.diagram.honeycomb}]{
  		\includegraphics[width=7cm]{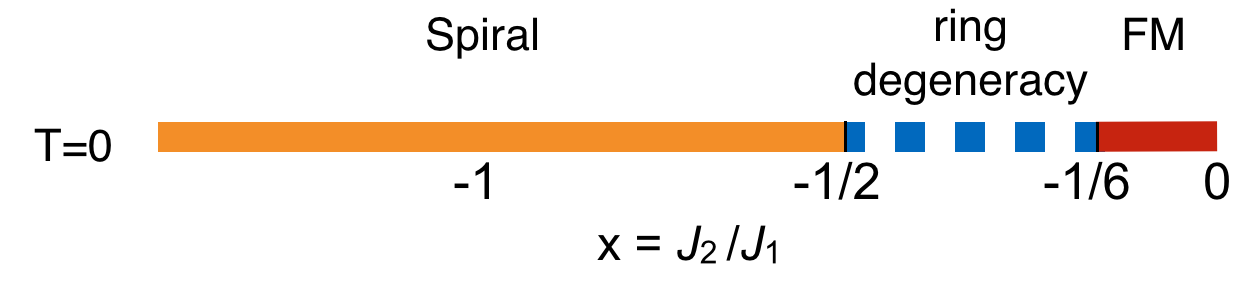}
	}
	\end{minipage}
	\caption{ 
	         {\bf Magnetic interactions in \CCO\ and mapping onto effective Honeycomb--lattice model.} 
		({\bf a}) Bilayer breathing Kagome (BBK) lattice of spin--1/2 Cr$^{5+}$ ions in \CCO.  
		Interactions are labelled following the conventions of Balz~{\it et al.}
		\cite{balz16-NatPhys12, balz17-PRB95} (cf. Table~\ref{table:exchange.interactions}).
		({\bf b}) Effective honeycomb lattice formed by spin--3/2 moments on FM plaquettes.  
		%
		({\bf c}) Finite--temperature phase diagram of the BBK model, Eq.~(\ref{eq:H.BBK}), 
		as determined by classical Monte Carlo simulation, 
		allowing $J_{\sf eff} \equiv J_{31} = J_{32}$ to vary, with all other parameters taken from \CCO.
                 Dotted lines correspond to peaks in the specific heat $C(T)$, and indicate the onset 
                 of correlations corresponding to spiral/ferromagnetic states at low temperature.
		White spots the show parameter--ratio and temperatures associated with 
		inelastic neutron--scattering experiments on \CCO\ \cite{balz16-NatPhys12, balz17-PRB95}. 
		{\bf Inset:} ``ring'' in equal--time structure factor $S({\bf q})$, 
		characteristic of the spiral spin liquid.
		({\bf d}) Classical ground state of the effective honeycomb--lattice model, 
		Eq.~(\ref{eq:H.HCM}), following \cite{rastelli79-PhysicaB+C97,fouet01-EPJB20}. 
		%
		The spiral spin liquid can be traced to a highly--degenerate manifold 
		of classical ground states occurring for $-1/2  < x  <  -1/6$.  
		}
	\label{fig:model.and.phase.diagram}
\end{figure*}

%
The first surprise in \CCO\ is a chemical one; the highly--unusual, spin--1/2,  
Cr$^{5+}$ valence state \cite{arcon98-JAmCeramSoc81}.  
Structurally, \CCO\ has much in common with the quantum dimer system 
SrCr$_2$O$_8$ \cite{quintero-castro10,wang16-PRL116}.
However the inclusion of non--magnetic Cr$^{6+}$ ions 
converts the triangular lattice of SrCr$_2$O$_8$ into a bilayers of a 
``breathing'' Kagome lattice \cite{balz17-JPCM29} 
--- cf. Fig.~\ref{fig:BBK.model}.   
These bilayers have a six--site unit cell, and very low symmetry (R3c) 
\cite{gyepesova13-ActaCrysC69,balz17-JPCM29}.  
%



The magnetic properties of \CCO\ are every bit as exotic as its 
chemistry.   
Curie--law fits to magnetic susceptibility suggest
predominantly ferromagnetic (FM) exchange interactions,  
with $\theta_{\sf CW} = 2.35\ \text{K}$ \cite{balz17-PRB95}. 
Meanwhile, the lack of magnetic anisotropy, coupled with the fact 
that \CCO\ is a good insulator, suggest that a Heisenberg model 
should provide a good starting point for understanding its magnetism 
\cite{balz16-NatPhys12,balz17-PRB95,balz17-JPCM29}.  
At that point, however, all semblance of conventional magnetism ceases.   
\CCO\ does not order down to the lowest temperatures measured, 
with persistent spin dynamics found in $\mu$SR down to 
$19\ \text{mK}$ \cite{balz16-NatPhys12}. 
%
%
Consistent with this, neutron scattering experiments find no magnetic 
Bragg peaks down to $90\ \text{mK}$ \cite{balz16-NatPhys12,balz17-PRB95}.
Instead, scattering is predominantly inelastic and highly--structured, 
with results at $0.25\ \text{meV}$ showing hints of a ring centered 
on (1,1,0) [Fig.~\ref{fig:Exp_025meV}], while scattering at $0.90\ \text{meV}$
suggests a bow--tie like structure centered on (0.5,\ 0.5,\ 0) 
[Fig.~\ref{fig:Exp_090meV}].
\begin{table}[H]
\caption{{\bf Exchange interactions in bilayer breathing--Kagome model 
of \CCO,  and mapping onto effective honeycomb--lattice model.}
Magnetic interactions within the spin--1/2 bilayer breathing--Kagome (BBK) model, 
${\cal H}_\text{\sf BBK}$ [Eq.~(\ref{eq:H.BBK})], are labelled following the 
convention of Balz~{\it et al.} \cite{balz16-NatPhys12,balz17-PRB95}, 
as illustrated in Fig.~\ref{fig:BBK.model}.
Experimental values are taken from fits to inelastic neutron scattering 
in applied magnetic field \cite{balz16-NatPhys12,balz17-PRB95}.   
The mapping onto an effective spin--3/2 honeycomb--lattice (HCM) model,  
${\cal H}_\text{\sf HCM}$ [Eq.~(\ref{eq:H.HCM})], is illustrated in Fig.~\ref{fig:BBK.model}. 
}
\begin{center}
\begin{tabular}{|c|c|c|}
	\hline
	\hspace{15pt} ${\cal H}_\text{\sf BBK}$ [Eq.~(\ref{eq:H.BBK})] 	
	& 	 \hspace{0.5cm}\CCO\ \cite{balz16-NatPhys12,balz17-PRB95}\hspace{0.5cm} 
	&       ${\cal H}_\text{\sf HCM}$ [Eq.~(\ref{eq:H.HCM})] \\ 
	\hline
	$J_\text{0}$ & $-0.08(4)\ \text{meV}$ & $J_\text{1}$ \\ 
	\hline
	$J_\text{21}$ & $-0.76(5)\ \text{meV}$ & --  \\ 
	\hline 
	$J_\text{22}$ & $-0.27(3)\ \text{meV}$ &	--  \\ 
	\hline 
	$J_\text{31}$ & $\phantom{-}0.09(2)\ \text{meV}$  &	$J_\text{2}$ \\ 
	\hline 
	$J_\text{32}$ & $\phantom{-}0.11(3)\ \text{meV}$ &	$J_\text{2}$ \\ 
	\hline 
\end{tabular} 
\end{center}
\label{table:exchange.interactions}
\end{table}


In contrast with SrCr$_2$O$_8$ 
\cite{quintero-castro10,wang16-PRL116}, the magnetization of 
\CCO\ rises rapidly in applied magnetic field, implying a gapless 
ground state \cite{balz17-PRB95}.
The magnetisation also saturates at the relatively low field of $13\ \text{T}$, 
making it possible to carry out inelastic neutron scattering (INS) experiments 
in the field--polarised state \cite{balz16-NatPhys12,balz17-PRB95}.
These reveal gapped, two--dimensional spin--wave excitations 
[Fig.~\ref{fig:spin.dynamics.11T}a], well--described by a Heisenberg 
model for single bilayer \cite{balz16-NatPhys12,balz17-PRB95}
\begin{equation}
	{\cal H}_{\sf BBK} 
	= \sum_{\langle i j \rangle} J_{ij} {\bf S}_i \cdot {\bf S}_j - {\bf B} \cdot \sum_i {\bf S}_i \; ,
    \label{eq:H.BBK}
\end{equation}
where ${\bf B}$ is the external magnetic field, and $J_{ij}$ the exchange interaction
on the first--neighbour bond $ij$ [cf. Table~\ref{table:exchange.interactions}].  
The strongest exchange interactions are FM, and occur within triangular 
plaquettes in alternating layers [Fig.~\ref{fig:BBK.model}].   
On reducing magnetic field, the sharp features associated with spin--waves 
retain their identity, but move down in energy, with the lowest--lying band 
vanishing into the elastic line below about $1\ \text{T}$  \cite{balz17-PRB95}.   
At the same time, heat capacity measurements show a systematic change 
at $1\ \text{T}$, consistent with the closing of a gap \cite{balz17-PRB95}.


While spin--liquid behaviour has been discussed in a wide range of
two--dimensional systems \cite{balents10-Nature464,anderson73-MatResBull8,lee08-Science321,shimizu03-PRL91,masutomi04-PRL92,han12-Nature492,kitaev06-AnnPhys321,jackeli09-PRL102,banerjee16-NatMater15,baek17-PRL119}, the origin of the gapless spin liquid in \CCO\ presents an 
entirely new challenge to theory.
In addition to experiment, Balz {\it et al.}~\cite{balz16-NatPhys12} 
present the results of a pseudo--fermion functional renormalisation group 
(PFFRG) analysis of ${\cal H}_\text{\sf BBK}$ [Eq.~(\ref{eq:H.BBK})].   
These calculations provide information about the ground--state properties 
of the model, including its static structure factor, $S({\bf q}, \omega=0)$.   
Tantalizingly, for parameters taken from experiment, PFFRG 
calculations are consistent with a spin-liquid ground state, and predict 
a ring of scattering in $S({\bf q}, \omega =0)$, centered on $(1,1,0)$ 
\cite{balz16-NatPhys12}.   
However, because of the involved nature of the calculations, and 
complexity of the model, the origin of this ring structure remains obscure.
And, for the time being at least, a direct comparison with INS results 
above the elastic line is out of reach.   
The challenge, then, is to explain how the seemingly--simple magnetism 
of \CCO\ in high magnetic field, evolves into an entirely new form 
of gapless spin liquid for fields of less than $1\ \text{T}$.   

\section*{Results}


To address the orgin of spin liquid behaviour in \CCO\, 
we have used semi--classical molecular dynamics 
(MD) simulation of ${\cal H}_\text{\sf BBK}$ [Eq.~(\ref{eq:H.BBK})] 
to explore how spin excitations evolve as function of magnetic field. 
Within this approach, spin configurations are drawn from a classical Monte Carlo 
simulation of ${\cal H}_\text{\sf BBK}$, carried out at finite temperature, 
and evolved according to the Heisenberg Equations of Motion 
\cite{moessner98-PRL80,conlon10-PRB81,taillefumier14-PRB90}.   
We concentrate on a single bilayer, with parameters taken from experiment 
[cf. Table~\ref{table:exchange.interactions}].


\begin{figure*}[t]
	\centering	
	\hspace{-1.5cm}
	\begin{minipage}[t]{0.97\textwidth}
	\centering
  	\includegraphics[height=4.5cm]{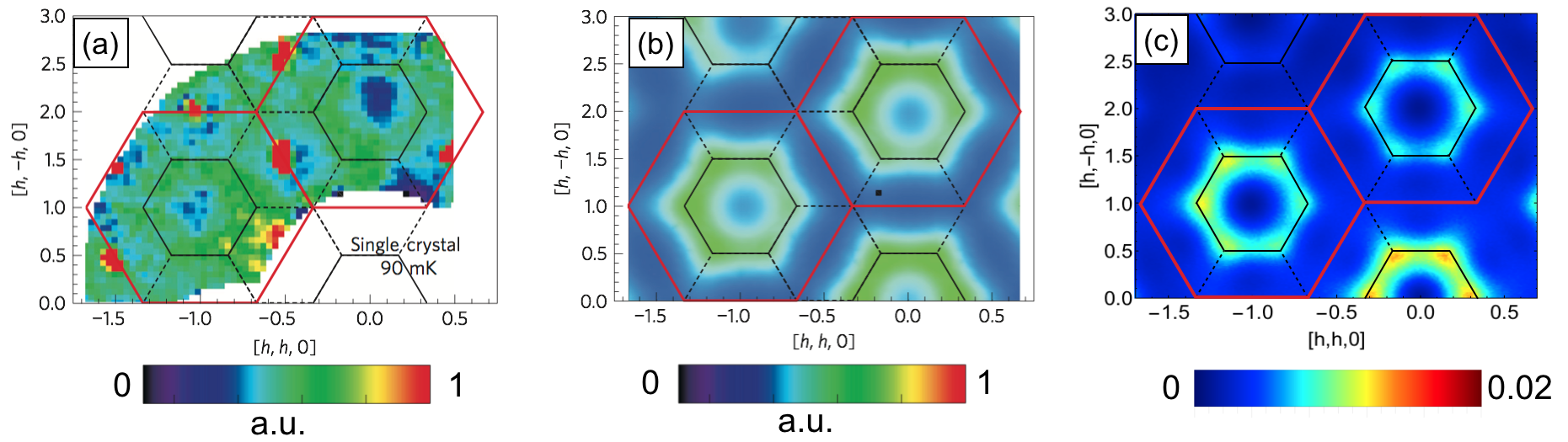}
         \end{minipage}
	\caption{ 
	         {\bf Comparison of ``ring'' structure in spin correlations, as revealed by 
	        experiment and simulation}.
	         ({\bf a}) Dynamical structure factor measured in inelastic neutron scattering (INS) 
	         experiments on \CCO\ at $T = 90\ \text{mK}$, $E = 0.25\ \text{meV}$, 
	         reproduced from \cite{balz16-NatPhys12}.
	         ({\bf b}) Pseudo--Fermion Functional Renormalisation Group (PFFRG) 
	         results for static structure factor $S({\bf q},\omega = 0)$, also reproduced 
	         from \cite{balz16-NatPhys12}, showing a ring--structure near to the 
	         Brillouin Zone (BZ) boundary.        
	         ({\bf c}) Equivalent Molecular Dynamics (MD) simulation results for   
	         $S({\bf q},\omega)$ at $T = 222\ \text{mK}$, 
	         $\hbar\omega = 0.23\ \text{meV}$,  showing the same ring structure. 
		}
	\label{fig:ring.structure}
\end{figure*}


The results of such a simulation, carried out for a temperature of $T = 222\ \text{mK}$, 
in the absence of magnetic field ($B = 0\ \text{T}$), are documented in the first 
Animation given in the Supplementary Information \cite{first-animation}.
As the Animation shows, the spins continue to fluctuate, even at this low 
temperature, and parallel calculations of the equal--time structure factor 
$S({\bf q})$ reveal a ``a ring'' structure, with no signs of long range order 
[cf. Inset to Fig.~\ref{fig:phase.diagram.BBK}].
%
%
On closer inspection, the simulation reveals that the spins exhibit both 
slow and fast dynamics, and that these have very different character, 
with a slow procession of locally collinear spins, mixing with fast 
fluctuations of seemingly--uncorrelated spins.
(For clarity, in the Animation, spins which rotate quickly have been 
coloured red).


All of these facts are consistent with the observation that \CCO\ enters
a spin--liquid state at low temperatures, in which the fluctuations on different 
timescales show qualitatively different character.
And the results of the MD simulations also stand up to a more quantitative 
comparison with experimental data.   
In Fig.~\ref{fig:EnergySlicesB_0T}, we 
show simulation results for the dynamical structure factor $S({\bf q}, \omega)$,
side--by--side with results from INS taken from \cite{balz16-NatPhys12}.
To facilitate comparison, simulation results have been convoluted with 
a Gaussian mimicking experimental resolution, and a magnetic 
form factor appropriate to a Cr$^{5+}$ ion.
Both experiment and theory show broad "bow--tie" like structures at 
intermediate to high energy, and rings of scattering at low energy.
(Bright ``spots'' seen in INS near to zone centers 
represent scattering from phonons, and do not form part 
of the magnetic signal \cite{balz16-NatPhys12}).  


Clearly, the MD simulations capture important elements of 
the physics of \CCO.   
Moreover, 
the ``ring'' found at low energies in MD simulation corresponds exactly 
to the ring found in the static structure factor $S({\bf q}, \omega=0)$ 
in PFFRG calculations \cite{balz16-NatPhys12} --- cf. Fig.~\ref{fig:ring.structure}.
The question which remains, is what does this tell us about the nature 
and origin of the spin--liquid state?

\begin{figure*}[t]
	\captionsetup[subfigure]{labelformat=empty, farskip=2pt,captionskip=1pt}
	\centering	
	\begin{minipage}[t]{0.47\textwidth}
		\centering
		\subfloat[]{
  			\includegraphics[width=0.95\columnwidth]{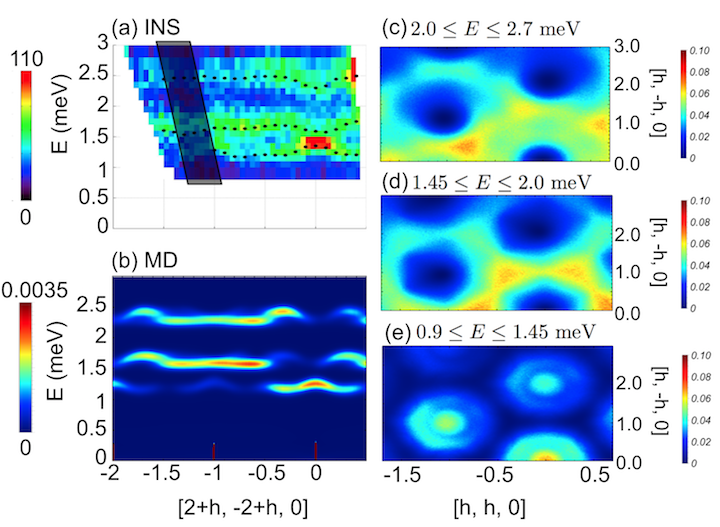}
		}
		\caption{
		{\bf Spin excitations of \CCO\ in high magnetic field ($B=11\ \text{T}$).}
		({\bf a}) Spin excitations measured in inelastic neutron scattering \cite{balz16-NatPhys12}, 
	        integrated over $\Delta{\bf q} = \pm 0.1\ \text{rlu}$ perpendicular to the cut in reciprocal space, 
	        showing dispersing excitations with three distinct energy scales.
		The dashed line shows a fit to linear spin wave (LSW) theory, 
		for parameters given in Table~\ref{table:exchange.interactions}.   
		({\bf b}) Equivalent results for the dynamical structure factor $S({\bf q},\omega)$, 
		taken from molecular dynamics (MD) simulations 
		of  ${\cal H}_\text{\sf BBK}$ [Eq.~(\ref{eq:H.BBK})], with parameters 
		taken from experiment [Table~\ref{table:exchange.interactions}].
		({\bf c})--({\bf e}) Contribution to the energy--integrated structure factor $S({\bf q})$
		coming from excitations 
		in the higher--, intermediate-- and lower--energy bands visible in ({\bf b}), 
		within MD simulation.   
		Bow--tie like features are clearly visible in the intermediate--energy bands, 
		shown in ({\bf d}).
		%
	         }
		\label{fig:spin.dynamics.11T}
	\end{minipage}
	%
	\quad \quad
	%
	%
	\captionsetup[subfigure]{labelformat=empty, farskip=2pt,captionskip=1pt}
	\centering	
	\begin{minipage}[t]{0.47\textwidth}
		\centering
		\subfloat[]{
  			\includegraphics[width=0.95\columnwidth]{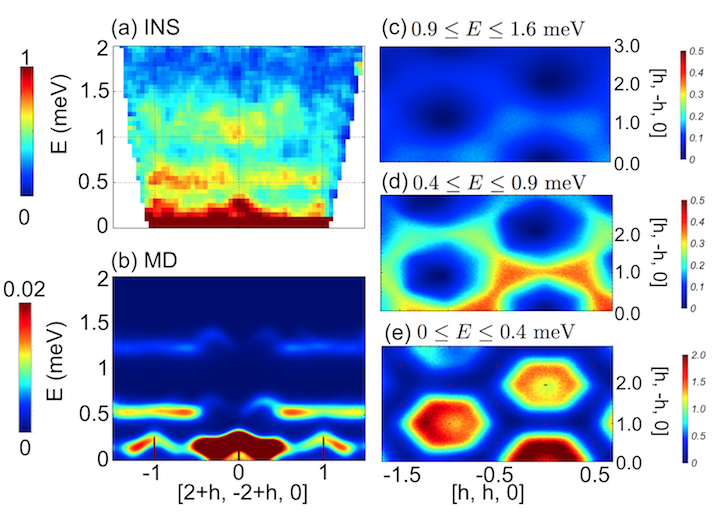}
		}
		\caption{
		{\bf Spin excitations of \CCO\ in intermediate magnetic field ($B=2\ \text{T}$).}
		({\bf a}) Spin excitations measured in inelastic neutron scattering \cite{balz16-NatPhys12}, 
	        integrated over $\Delta{\bf q} = \pm 0.2\ \text{rlu}$ perpendicular to the cut in reciprocal space.
		({\bf b}) Equivalent results 
		from molecular dynamics (MD) simulations 
		of  ${\cal H}_\text{\sf BBK}$ [Eq.~(\ref{eq:H.BBK})], with parameters 
		taken from experiment [Table~\ref{table:exchange.interactions}].
		({\bf c})--({\bf e}) Contribution to the energy--integrated structure factor $S({\bf q})$
		coming from excitations 
		in the higher--, intermediate-- and lower--energy bands visible in ({\bf b}), 
		within MD simulation.   
		Bow--tie like features are clearly visible in the intermediate--energy 
		bands, shown in~({\bf d}).
		Ring--like features are starting to form in the lower--energy bands, 
		shown in~({\bf e}).
		}
		\label{fig:spin.dynamics.2T}
	\end{minipage}
\end{figure*}

To gain more insight into this question, we have used MD simulation 
to track the evolution of spin dynamics from the saturated state at 
high magnetic field, into the spin liquid at $B = 0\ \text{T}$.   
In Fig.~\ref{fig:spin.dynamics.11T} we show a comparison of INS results 
and MD simulation for a magnetic field of $B = 11\ \text{T}$.
Parameters for the MD simulation have been taken from experiment 
[cf. Table~\ref{table:exchange.interactions}] and, following \cite{balz17-PRB95}, 
results have been convoluted with a Gaussian of 
$\text{FWHM} = 0.2\ \text{meV}$, mimicking finite experimental resolution.
The agreement between simulation and experiment is excellent.


Both experiment and simulation show excitations on three different 
energy scales; each of these corresponds to two bands of 
nearly--degenerate, transverse, spin--wave excitations \cite{balz17-PRB95}.   
The excitations at higher and intermediate energy are similar, but 
those at lower energy have a very different character, and contribute 
in different ways to the equal--time structure factor $S({\bf q})$.
In Fig.~\ref{fig:spin.dynamics.11T}c, we show the results 
for the dynamical structure factor $S({\bf q}, \omega)$, integrated
over a range of energies corresponding the two highest--energy 
spin--wave modes.
Equivalent results for the intermediate--energy excitations 
are shown in Fig.~\ref{fig:spin.dynamics.11T}d.
Both exhibit ``bow--tie'' structures ---  e.g. centred on ${\bf q} = (0.5, 0.5, 0)$ --- 
which link to form a broad network of scattering within the 
$[h,l,0]$ plane.
Meanwhile, equivalent calculations for the low--energy 
spin excitations show that their spectral weight 
is concentrated within the ``holes'' in this network 
[Fig.~\ref{fig:spin.dynamics.11T}e].


\begin{figure*}[t]
	\captionsetup[subfigure]{labelformat=empty, farskip=2pt,captionskip=1pt}
	\centering	
	\begin{minipage}[t]{0.47\textwidth}
		\centering
		\subfloat[]{
  			\includegraphics[width=0.95\columnwidth]{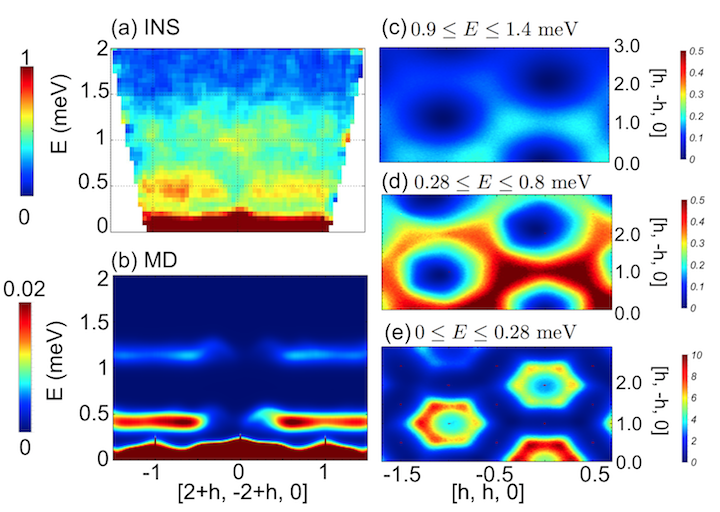}
		}
		\caption{
		{\bf Spin excitations of \CCO\ at the critical value of magnetic field} ($B=1\ \text{T}$).
		({\bf a}) Spin excitations measured in inelastic neutron scattering \cite{balz16-NatPhys12}, 
                 integrated over $\Delta{\bf q} = \pm 0.2\ \text{rlu}$ perpendicular to the cut in reciprocal space.
		({\bf b}) Equivalent results 
		from molecular dynamics (MD) simulations 
		of  ${\cal H}_\text{\sf BBK}$ [Eq.~(\ref{eq:H.BBK})], with parameters 
		taken from experiment [Table~\ref{table:exchange.interactions}].
		({\bf c})--({\bf e}) Contribution to the energy--integrated structure factor $S({\bf q})$
		coming from excitations 
		in the higher--, intermediate-- and lower--energy bands visible in ({\bf b}), 
		within MD simulation.   
		The gap to spin transverse spin excitations has closed, marking 
		the onset of the low--field spin--liquid regime.
		This is characterised by a well--formed ring of correlations in 
		$S({\bf q})$, coming from spin excitations with low energy.
		%
	         }
		\label{fig:spin.dynamics.1T}
	\end{minipage}
	%
	\quad \quad
	%
	%
	\captionsetup[subfigure]{labelformat=empty, farskip=2pt,captionskip=1pt}
	\centering	
	\begin{minipage}[t]{0.47\textwidth}
		\centering
		\subfloat[]{
  			\includegraphics[width=0.95\columnwidth]{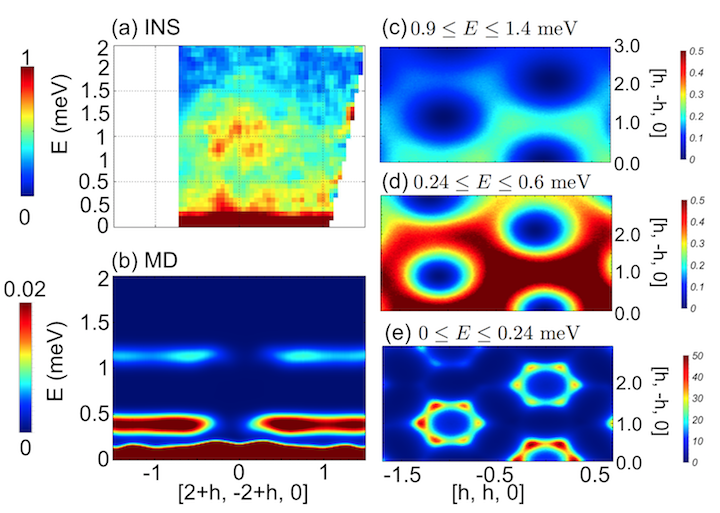}
		}
		\caption{
		{\bf Spin excitations of \CCO\ in the absence of magnetic field ($B=0\ \text{T}$).}
		({\bf a}) Spin excitations measured in inelastic neutron scattering \cite{balz16-NatPhys12}, 
	         integrated over $\Delta{\bf q} = \pm 0.2\ \text{rlu}$ perpendicular to the cut in reciprocal space.
		({\bf b}) Equivalent results 
		from molecular dynamics (MD) simulations 
		of  ${\cal H}_\text{\sf BBK}$ [Eq.~(\ref{eq:H.BBK})], with parameters 
		taken from experiment [Table~\ref{table:exchange.interactions}].
		({\bf c})--({\bf e}) Contribution to the energy--integrated structure factor $S({\bf q})$
		coming from excitations 
		in the higher--, intermediate-- and lower--energy bands visible in ({\bf b}), 
		within MD simulation.   
%
		}
		\label{fig:spin.dynamics.0T}
	\end{minipage}
\end{figure*}

As magnetic field is reduced, the gap to each band of 
excitations reduces, as illustrated in Fig.~\ref{fig:spin.dynamics.2T}.
At the same time, the individual character of spin excitations 
at different energy scales becomes easier to resolve; ``bow--tie''
patterns become more clearly marked at higher energies 
[Fig.~\ref{fig:spin.dynamics.2T}{\bf c} and Fig.~\ref{fig:spin.dynamics.2T}c], 
and low--energy fluctuations start to form a ring at the zone boundary 
[Fig.~\ref{fig:spin.dynamics.2T}e].
For this value of field, a large part of the spectral 
weight in experiment is obscured by quasi--elastic scattering, 
and it is hard to estimate whether the gap to transverse spin 
excitations is open or closed [Fig.~\ref{fig:spin.dynamics.2T}a].
However in simulation, it is clear that the gap remains open 
[Fig.~\ref{fig:spin.dynamics.2T}b --- see also Fig.~S1d5 of 
Supplementary Information].
The gap to transverse excitations finally closes at $B = 1\ \text{T}$, 
as shown in Fig.~\ref{fig:spin.dynamics.1T}, marking the onset 
of the low--field spin liquid phase.
The critical value of field obtained from simulation, 
$B = 1.0(5)\ \text{T}$, is in good correspondence with thermodynamic 
measurements, which suggest the onset of gapless spin--liquid 
behaviour at  $B = 1\ \text{T}$ \cite{balz16-NatPhys12}.  


We are now in position to address a crucial question, so far as the nature 
of the spin liquid is concerned:
what is the nature of the spin correlations at the point at which this gap
closes?
We address first the ``ring'' seen in the lowest--energy 
excitations in MD simulation, and in the PFFRG calculations at 
$\omega= 0$.
Ring--degeneracies of this form are familiar from a range of
models which 
display spin--liquid ground states 
\cite{rastelli79-PhysicaB+C97,fouet01-EPJB20,okumua10-JPSJ79,seabra16-PRB93}.
Classically, they can be understood as an enesmble of degenerate spirals, 
and have been dubbed ``spiral spin liquids'' \cite{bergman07-NatPhys7}.
In the present case, we can also easily understand the origin of the 
``ring'' degeneracy.
The strongest exchange interactions in \CCO\ are FM and occur on 
alternating plaquettes [cf. Fig.~\ref{fig:BBK.model}]. 
At low energies, these FM plaquettes behave like a honeycomb lattice 
of effective spin--3/2 moments  [cf. Fig.~\ref{fig:HCM.model}]
\begin{equation}
{\cal H}_{\sf HCM}
= J_1 \sum_{\langle ij \rangle_1} \ {\bf S}_i \cdot {\bf S}_j 
+ J_2 \sum_{\langle ij \rangle_2}  {\bf S}_i \cdot {\bf S}_j \; ,
\label{eq:H.HCM}
\end{equation}
with FM $J_1$ and AF $J_2$ [cf.~Table~\ref{table:exchange.interactions}].
For \mbox{$-1/2 \leq J_2/J_1 \leq -1/6$},  the classical ground state 
of ${\cal H}_{\sf HCM}$ is known to be a degenerate set of spirals with 
wave vectors belonging to a ring \cite{rastelli79-PhysicaB+C97}, 
providing all of the ingredients needed to form a classical 
spin liquid 
\cite{seabra16-PRB93}.
And a quantum spin liquid, occurring for a somewhat broader
range of parameters, is indicated in exact--diagonalisation studies 
of the corresponding spin--1/2 model \cite{fouet01-EPJB20}.  


Experimental estimates for \CCO\ yield 
\mbox{$-3.1 \lessapprox J_2/J_1 \lessapprox -0.6$} [Table~\ref{table:exchange.interactions}], 
neighbouring the degenerate regime, but within a region of parameter 
space where the classical ground state of ${\cal H}_{\sf HCM}$ is an ordered 
spiral 
[Fig.~\ref{fig:phase.diagram.honeycomb}].  
In keeping with this, our MC simulations show a strong anomaly in heat 
capacity at $T \approx 60\ \text{mK}$, consistent with a tendency 
to order.  
However, above this temperature, the entropy associated with 
ring--degeneracy predominates, and classical MC simulations 
find a spiral--spin liquid, characterised by a ring in 
$S({\bf q})$ 
[Fig.~\ref{fig:phase.diagram.honeycomb}].
And, given that line--degeneracies lead to logarithmic  
divergences in spin--wave corrections in two dimensions 
\cite{chandra88-PRB38,smerald10}, we anticipate that any 
classically--ordered ground state would quickly be 
eliminated by quantum fluctuations. 




We now turn to the ``bow--tie'' features observed at finite energy.
Structure of this form is already familiar from the Kagome--lattice 
antiferromagnet, where they have the interpretation of 
``pinch points'', singular points in scattering associated with a local 
constraint on spin configurations 
\cite{garanin99-PRB59,zhitomirsky08-PRB78,taillefumier14-PRB90}.
As such, they are a signal feature of an entirely different kind 
of spin liquid --- a Coulombic phase with an emergent $U(1)$ gauge 
structure \cite{henley10}.
The pinch points seen in simulations of \CCO\ 
[cf. Fig.~\ref{fig:spin.dynamics.11T}] are encoded in bands of 
transverse spin excitations.
These bands occur in pairs and, at lower value of magnetic 
field, overlap with much broader bands of longitudinal 
excitations, documented in the Supplementary Information.
It is possible to understand these pinch points 
through 
the dynamical stability of a local constraint, 
a theme which will be explored further elsewhere \cite{han-unpub}.
Meanwhile, the continuous evolution of the ``bow--tie'' structure from 
$B = 11\ \text{T}$ \mbox{[Fig.~\ref{fig:spin.dynamics.11T}c,d]}  
to $B = 0\ \text{T}$ \mbox{[Fig.~\ref{fig:spin.dynamics.0T}c,d]}, 
suggests that 
this local constraint 
also has a role to play in the 
fluctuations of the spin liquid.    


Armed with these new insights, we are finally in a position to return to 
simulations carried out in the absence of magnetic field, and interpret 
the very different structure seen at different timescales. 
In the Second Animation given in the Supplementary Information 
\cite{second-animation}, we exhibit the same results as shown in the 
First Animation, but resolved into the energy--bands of the slow, 
intermediate and fast fluctuations.  


The picture which emerges is of a ``three--speed'' spin liquid.
The slowest fluctuations (panel on left) exhibit the behaviour 
expected of the ``spiral spin liquid'', with spins ferromagnetically 
aligned on the individual triangular plaquettes of the lattice, and 
exhibiting slow, collective rotations.
Meanwhile the intermediate fluctuations (central panel), and fast 
fluctuations (panel on right), show a very different character.
Spins are antiferromagnetically correlated on triangular plaquettes, 
and overlaid with a fast, correlated, single spin--flip dynamics 
redolent of the ``weather vane'' modes in Kagome 
lattice--antiferromagnets \cite{harris92-PRB45}.   
Fluctuations on different timescales can best be compared by 
adjusting the frame rate so as to approximately match 
the speeds of fluctuations in each panel (second half of the 
Second Animation \cite{second-animation}).
Viewed in this way, we obtain a real--space picture of the very 
different correlations resolved in the partially--integrated 
structure factors shown in Fig.\ref{fig:spin.dynamics.0T}c,d,   
and in Fig.\ref{fig:spin.dynamics.0T}e.  

\section*{Discussion}

The semi--classical simulations described in this Article provide a good
account of many of the features observed in experiments on \CCO, and 
paint a colourful picture of a spin liquid existing on multiple timescales.
Just how unusual is this?
Everyday life teaches that ordinary substances can have very different 
properties, depending on how long you wait.   
Water, the liquid which holds the key to life, acts 
like a solid in sudden impacts.    
Ice, meanwhile, is a solid which flows like a liquid over glacial timescales.
Quantum systems too, hold many such surprises.
In the quasi--one dimensional, quantum spin--ladder CaCu$_2$O$_3$, static magnetic 
order coexists with high--energy spinon excitations \cite{lake10-NatPhys6}.
Meanwhile, in the three--dimensional frustrated magnet Nd$_2$Zr$_2$O$_7$, 
all--in, all--out order coexists with excitations which support spin--ice like pinch points 
\cite{petit16-NatPhys12,benton16-PRB94}, an effect which has been dubbed 
``moment fragmentation'' \cite{brooks-bartlett14-PRX4}.
\CCO\ appears to be unique, however, in having properties which resemble
one spin liquid at low energies, and a different kind of spin liquid at higher energies.



This separation of timescales is relatively easy to visualise at a 
semi--classical level (cf. Second Animation \cite{second-animation}).
However, this level of approximation neglects all entanglement
between spins. 
To the best of our knowledge, this type of dynamics 
have yet to be explored for an entangled, fully--quantum system.  
So an important ``next step'' in the understanding of \CCO\ will
be more, explicitly quantum, analysis of ${\cal H}_\text{\sf BBK}$ 
[Eq.~(\ref{eq:H.BBK})].


In this respect, there are a great many interesting, open, questions. 
We have argued that the key to understanding the 
low--energy properties of \CCO\ is the closing of a gap to a 
(quasi--)degenerate ring of spin excitations at $B = 1\ \text{T}$.   
%
A scenario for such a phase transition has been previously discussed 
by Sedrakyan and coauthors, who dubbed the degenerate ring of 
bosonic excitations a ``moat--band'' \cite{sedrakyan14-PRB89}.
Considering the specific example of an XY model on the honeycomb 
lattice, with competing AF first-- and second--neighbour interactions, 
they argued the ground state should be a chiral spin liquid, breaking
inversion as well as time--reversal symmetries \cite{sedrakyan15-PRL114}.
In this picture, excitations have the character of Fermions, 
and it could be interesting to try to model the low--energy 
properties of \CCO\ in terms of Fermionic quasiparticles.   
The existence of (Fermionic) fractional spin--excitations is also 
one possible explanation for a feature of the INS data which 
is not well--captured by semi--classical simulations, namely a broad 
continuum of ``background'' scattering extending up to 
$\sim 1.5\ \text{meV}$ for $B=0\ \text{T}$ [cf. Fig.~\ref{fig:spin.dynamics.0T}].


The high--energy and high--field properties of \CCO\ also pose interesting 
new questions for theory.
The presence of competing FM and AF interactions immediately
raises the possibility of magnons forming bound states, leading to 
a phase with hidden, multipolar order \cite{chubukov91,shannon06,momoi06}.
And the longitudinal fluctuations found in MD simulations (see 
Supplementary Information) provide at least a hint that spin waves
are not the only excitations for $B > 1\ \text{T}$.
The presence of Kagome--like physics at high energy is also 
worth exploring in a quantum model, given the good agreement
between semi--classical simulation and experiment.
The pinch--point structures found at high energy, in high magnetic field, 
also have an interesting story to tell, with close parallels to work 
on Nd$_2$Zr$_2$O$_7$ \cite{benton16-PRB94}.  
An analytic theory of this phenonomen will be developed 
elsewhere \cite{han-unpub}.


Further experiment would also be very helpful in unravelling the 
mysteries of \CCO.
The simulations described in this Article makes explicit predictions for the 
structure, and magnetic--field evolution, of inelastic neutron scattering
which remain to be tested.
Here polarisation--analysis could be a vital new ingredient, both
to separate the magnetic signal from incoherent background in zero
magnetic field, and to resolve the different transverse and longitudinal
excitations in magnetic field [cf. Fig.~2, Supplemental Information].
Thermodynamic and transport measurements of \CCO, 
which probe the low--energy excitations of its QSL ground
state, could also be revealing.   
And given that \CCO\ may be proximate to an AF Kagome--like 
pump--probe experiments, for example
using fast optics,  
might be particularly interesting.


Finally, since completing this work, we have become aware of a 
parallel study of \CCO, 
by Biswas and Damle  \cite{biswas-arXiv}.
Their preprint provides a complementary analysis of the low--energy 
properties of ${\cal H}_\text{\sf HCM}$ [Eq.~(\ref{eq:H.HCM})], in the absence 
of magnetic field, and specifically addresses a question which we have 
not attempted to answer here, namely the fate of  classical spins at 
temperatures \text{$T \lesssim 60\ \text{mK}$}. 

\section*{Conclusions}

\CCO\ is a remarkable magnet, in which spin--1/2 Cr$^{5+}$ ions form 
a bilayer breathing--kagome (BBK) lattice with complex, 
competing exchange interactions \cite{balz17-JPCM29}.  
A combination of heat--capacity, magnetization, $\mu$SR, neutron--scattering, 
and AC susceptibility experiments reveal \CCO\ to be a gapless quantum 
spin liquid (QSL), showing no sign of magnetic order down to $19\ \text{mK}$ 
\cite{balz16-NatPhys12,balz17-PRB95}.
This spin liquid is charcaterised by spin fluctuations which 
show qualitatively different character on different timescales.


To better understand the nature and origin of the spin liquid in \CCO, we have 
carried out large--scale semi--classical molecular--dynamics (MD) simulations of the 
minimal model of \CCO, a Heisenberg model on the BBK lattice, with 
parameters taken from experiment \cite{balz16-NatPhys12,balz17-PRB95}.
%
These simulations reveal a state where spins continue to fluctuate at 
very low temperatures, but the character of these 
fluctuations depends strongly on the timescale on which the dynamics 
are resolved, as shown in the 
Animations \cite{first-animation,second-animation}.   


We identify fluctuations at low energies with a ``spiral spin 
liquid'', characterised by a ring of scattering in ${\bf q}$--space, 
and formed when the gap to a (quasi--)degenerate set of excitations 
closes at $B = 1\ \text{T}$ [cf. Fig.~\ref{fig:ring.structure}].
This spiral spin liquid can be described by an effective 
spin--3/2 Heisenberg model on a honeycomb lattice, formed
by spin--1/2 moments on the triangular plaquettes of 
the BBK lattice in \CCO\ [cf. Fig.~\ref{fig:BBK.model}, Fig.~\ref{fig:HCM.model}].
The FM correlations of spins within these plaquettes 
are evident in the collective motion resolved at low energy 
in MD simulation (cf. Second Animation~\cite{second-animation}). 


Meanwhile, fluctuations at higher energy inherit their character from 
the kagome--lattice antiferromagnet, and for \mbox{$B > 1\ \text{T}$}, 
are characterised by sharp pinch--points in scattering 
[Fig.~\ref{fig:spin.dynamics.11T}, Fig.~\ref{fig:spin.dynamics.2T}].
These pinch points are encoded in spin fluctuations transverse 
to the applied magnetic field [cf. Fig.~S2 of Supplemental Information], 
and can be resolved in MD simulations as collective rotations of
antiferromagnetically correlated spins on shorter timescales 
(cf. Second Animation \cite{second-animation}). 
When the gap to transverse spin excitations closes, at $B = 1\ \text{T}$, 
pinch points merge with excitations in the longitudinal channel
to give rise to the broader ``bow--tie'' features observed in inelastic 
neutron scattering at higher energy 
[cf. Fig.~\ref{fig:EnergySlicesB_0T}].



These simulations capture many of the features of \CCO; correctly reproducing 
the value of the critical field, $B = 1\ \text{T}$ \cite{balz16-NatPhys12,balz17-PRB95}; 
providing insight into the different structures seen in inelastic neutron scattering 
\cite{balz16-NatPhys12,balz17-PRB95}; 
and resolving the origin of the ring features found in Pseudo--Fermion Functional 
Renormalisation group (PFFRG) calculations \cite{balz16-NatPhys12}.
To the best of our knowledge, they also provide the first theoretical example of 
a system which behaves like different types of spin liquid on different timescales.


Given this disparity of behaviour, it is tempting to ask just how many spin 
liquids there are in \CCO?
Since a quantum system should have one, unique, ground state, 
at low temperature the answer to this 
question must, ultimately, be: ``{\it one}''.
None the less, the success of semi--classical simulations in describing 
experiment 
suggests that this one ground state must incorporate 
two different types of correlations; one described by effective spin--3/2 moments 
on a honeycomb lattice; and one corresponding to antiferromagnetic fluctuations of 
individual spin--1/2 moments on a bilayer breathing--Kagome lattice.
Unraveling the properties of this single, massively--entangled QSL,    
represents an exciting challenge for theory and experiment alike.

\section*{References}

\bibliography{paper}

\section*{Acknowledgements}

The authors are pleased to acknowledge helpful conversations with Owen Benton, 
Ludovic Jaubert and Mathieu Taillefumier, and are indebted to Christian Balz 
and Bella Lake for extended discussions and sharing information 
about experiments on \CCO.
The authors also gratefully acknowledge help with data visualization, provided by 
Pavel Puchenkov, of the Scientific Computing and Data Analysis Section, OIST.   
This work was supported by the Theory of Quantum Matter Unit, OIST.
Numerical calculations we carried out using HPC Facilities provided by OIST.

\section*{Author Contributions}

Both RP and HY contributed equally to this work and have 
been listed in alphabetical order.
RP carried out all numerical simulations, and analysis of numerical and 
experimental data.
HY initiated the project and carried out parallel analytic calculations. 
NS supervised the project, and prepared the manuscript, with 
input from HY and RP.

%

\section*{Methods}


{\bf Monte Carlo Simulation:}
All of the results presented in this Article are based on spin configurations drawn from 
classical Monte Carlo simulations of ${\cal H}_\text{\sf BBK}$ [Eq.~(\ref{eq:H.BBK})].
Monte Carlo simulations  were performed by using a local heat--bath 
algorithm \cite{Olive1986, Miyatake1986}, in combination with parallel 
tempering \cite{Swendsen1986, Earl2005}, and over--relaxation 
techniques \cite{Creutz1987}. 
A single MC step consists of $N$ local heat--bath updates on randomly chosen sites, 
and two over--relaxation steps, each comprising a \mbox{$\pi$--rotation} of all the spins in the lattice 
about their local exchange fields.
Simulations were performed in parallel for replicas at $200$ different 
temperatures, with replica--exchange 
initiated by the parallel tempering algorithm every $10^2$ MC steps.
Results for thermodynamic quantities were averaged over $10^6$ statistically independent  
samples, after initial $10^6$ MC steps for simulated annealing and $10^6$ MC steps for 
thermalisation.
All presented results have been calculated for clusters of \mbox{$N = 13824$} sites.


{\bf Molecular Dynamics (MD) Simulation:}
Our MD simulations are based on the 
semi--classical, Heisenberg equations of motion 
\begin{eqnarray}
	\frac{d {\bf S}_i}{d t} 	= \frac{\mathrm{i}}{\hbar} \big[ {\cal H}_{\sf BBK}, {\bf S}_i \big]   
					= \bigg( \sum_j J_{ij} {\bf S}_j - {\bf B} \bigg) \times {\bf S}_i		\; ,
\label{eq:E.of.M}
\end{eqnarray}
where $j$ accounts for all nearest--neighbouring sites of $i$ and $J_{ij}$ is given in 
Table~\ref{table:exchange.interactions}.
Numerical integration of Eq.~(\ref{eq:E.of.M}) was carried out 
using a 4$^{th}$ order Runge--Kutta algorithm, as described in 
\cite{NumericalRecipes2007, OrdinaryDiffEquations1}.   
Spin configurations for MD simulation were taken from the thermal 
ensemble generated by classical MC simulations of ${\cal H}_{\sf BBK}$ 
at \mbox{$T=222\ \text{mK}$}, for parameters taken from experiment 
(cf. Table~\ref{table:exchange.interactions}).
The dynamical structure factor 
\begin{equation}
	S({\bf q}, \omega) = \frac{1}{\sqrt{N_t} N}
					\sum_{i, j}^N  \mathrm{e}^{ \mathrm{i} {\bf q} ({\bf r}_{i} - {\bf r}_j ) } 
					\sum_n^{N_t}  \mathrm{e}^{ \mathrm{i}  \omega \ n \delta t } \: 
					\langle {\bf S}_{i} (0) \cdot {\bf S}_{j} (t) \rangle   \; ,	
\label{eq:S.q.omega}
\end{equation}
was calculate using Fast Fourier Transform (FFT) \cite{FFTW05}, and averaged over spin dynamics 
obtained from $1000$ independent initial spin configurations.
MD simulations were performed for $N_t = 600$ time steps, with the time--increment $\delta t$ of 
\begin{equation}
	\delta t = \frac{t_{\text{max}}}{N_{t}} = \frac{2 \pi}{\omega_{\text{max}} }
\end{equation}
and a maximally resolvable frequency limit of \mbox{$\omega_{\text{max}} = 6\ \text{meV}$}.
To avoid numerical artefacts (Gibbs phenomenon \cite{MathimaticalPhyics}) coming from 
discontinuities of the finite time--window, at \mbox{$t = 0$} and \mbox{$t = t_{\text{max}}$}, the time 
sequence of spin configurations has been multiplied by a Gaussian 
envelop, imposing a Gaussian energy resolution of \mbox{FWHM $\approx$ 0.03\ meV}, 
on the numerically--obtained $S({\bf q}, \omega)$ [see Supplementary Information 
for further details].




{\bf Comparison with Experiment:}
Predictions for inelastic neutron scattering are plotted as 
\begin{eqnarray}
	\frac{d^2 \sigma}{d\Omega d E_f} 
					&\propto& I({\bf q}, \omega)
\end{eqnarray}
where we calculate 					
\begin{eqnarray}
	 I({\bf q}, \omega) &=& \mathcal{F}({\bf q})^2  \sum_{\alpha, \beta}  
					\bigg( \delta_{\alpha \beta} - \frac{q_{\alpha}q_{\beta}}{{\bf q}^2} \bigg)
					S^{\alpha \beta}({\bf q}, \omega)   \; .
\end{eqnarray}
Here $\mathcal{F}({\bf q})$ is the atomic form factor appropriate to a Cr$^{5+}$ ion, 
and following \cite{NeutronDataBooklet}, we write 
\begin{equation}
	\mathcal{F}({\bf q}) 
	=   \langle j_0 ({\bf q}) \rangle 
	+ \bigg(1 - \frac{2}{g} \bigg) \langle j_2 ({\bf q}) \rangle	\; .
\end{equation}
We consider gyromagnetic ratio $g = 2$, implying that 
$\langle j_2 ({\bf q}) \rangle$ plays no role.
The remaining function, $\langle j_0 ({\bf q}) \rangle $ can be parameterised as
\begin{equation}
	\langle j_0 ({\bf q}) \rangle	
	= 	A e^{- a ( |{\bf q} | / 4 \pi)^2} + B e^{- b ( | {\bf q} | / 4 \pi)^2} + C e^{- c (| {\bf q} | / 4 \pi)^2} + D	\; .
	\label{eq:FormFactor_j0}
\end{equation}
where, be consistent with earlier work \cite{balz-private}, 
coefficients are taken to be 
\begin{align}
	A &= -0.2602	\; ,	\  B = 0.33655	\; ,	\  C = 0.90596	\; , \ 	D = 0.0159	\\
	a &= 0.03958	\; ,	\  b = 15.24915	\; ,	\   c = 3.2568	\; .
\end{align}
For comparison with experiment, following \cite{balz16-NatPhys12,balz17-PRB95}, 
MD results for $S({\bf q}, \omega)$ have further been convoluted in energy with 
a Gaussian of \mbox{FWHM = 0.2\ meV}.

\clearpage
\widetext


\section*{Supplementary material}

# Supplementary Material : How many spin liquids are there in $Ca_{10}Cr_7O_{28}$?

Rico Pohle,[1,][*] Han Yan,[1,][†] and Nic Shannon[1]

[1]*Okinawa Institute of Science and Technology Graduate University, Onna-son, Okinawa 904-0412, Japan*

(Dated: November 10, 2017)

[*] rico.pohle@oist.jp
[†] han.yan@oist.jp

# ANIMATIONS OF MOLECULAR DYNAMICS SIMULATIONS

### First Animation

In the First Animation, we show a "fly through" of spins on the breathing bilayer Kagome (BBK) lattice, with time evolution taken from a single molecular dynamics (MD) simulation of $\mathcal{H}_{\sf BBK}$ [Eq. (1)], for parameters relevant to $Ca_{10}Cr_7O_{28}$ [Table 1 of the main text]. The simulation was carried out for a cluster of $N = 5400$ sites, at a temperature of $T = 222$ mK. The total number of time steps in the simulation was $N_t = 6000$, and in order to obtain a smooth rotation of spins in the animation, the time step for each frame has been set to $\delta t \approx 0.1 \ \hbar^{-1}\text{meV}^{-1}$ [cf. Eq.(5) of the main text]. The animation was prepared using the open–source software package "Blender" [1]. To emphasize the dynamics on different timescales, spins have been color–coded according to their speed of rotation, with red indicating fast rotation, and green denoting slow rotation.

### Second Animation

In the Second Animation we show results taken from a single MD simulation of $\mathcal{H}_{\sf BBK}$, equivalent to that shown in the First Animation. However in this case, the time sequence for the spin dynamics has been separated into slow, intermediate and fast components, to emphasise the dynamics on different timescales. This is accomplished by first performing a fast Fourier transform (FFT) on the time sequence of each spin, then filtering the resulting signal in frequency space, using a digital analogue of a "band–pass" filter. The frequencies used for this band–pass filter are 0.750–1.500 meV (fast fluctuations); 0.225–0.750 meV (intermediate fluctuations]); and 0.00–0.225 meV (slow fluctuations). After filtering in frequency, a second FFT is used to reconstruct separate time sequences for slow, intermediate and fast fluctuations. The final result for each of these is presented in the three panels of the Second Animation.

In the second part of the Second Animation, the speed of playback for three time sequences has been adjusted, so as to match the characteristic speed of the relevant fluctuations. To accommodate this, a much longer sequence of $N_t = 130000$ time steps has been generated from MD simulation, using a much shorter time increment of $\delta t \approx 0.035 \ \hbar^{-1}\text{meV}^{-1}$. The adjustment of playback speed has been accomplished within "Blender" [1], by adjusting the number of frames included for each time sequence. Viewed in this way, the very different dynamics on different timescales is self–evident. Intermediate and fast fluctuations show dynamics redolent of the Heisenberg antiferromagnet on the Kagome–lattice. Meanwhile, qualitatively different dynamics, corresponding to the "spiral spin liquid", are resolved in the slow fluctuations of the spins. In all three cases, the dynamics is highly–localised, consistent with the weakly–dispersing bands of excitations resolved in MD results for $S(\mathbf{q}, \omega)$ [Fig. S1].

# EVOLUTION OF SPIN DYNAMICS IN APPLIED MAGNETIC FIELD

In Fig. S1 we compare the field–evolution of spin excitations found in inelastic neutron scattering (INS) measurements of $Ca_{10}Cr_7O_{28}$ [2] with MD simulations of $\mathcal{H}_{\sf BBK}$ [Eq. (1) of the main text]. Results for experiment are shown in the first row of panels, Fig. S1a. Both experiment and simulations have been carried out at a temperature $T \approx 220$ mK.

So far as simulations are concerned, our main tool for understanding spin fluctuations is the dynamical structure factor $S(\mathbf{q}, \omega)$, defined in Eq. 4 of the main text. In the second row of the figure [Fig. S1b], we show predictions for experiment derived from $S(\mathbf{q}, \omega)$; following [2], for comparison with experiment, results have been convoluted with a Gaussian of FWHM = 0.2 meV, and weighted by the atomic form factor for $Cr^{5+}$, as described in the Methods section of the main text. These results provide an excellent account of all of the sharp features seen in experiment [Fig. S1a].

For comparison, in the third row [Fig. S1c], we show "raw" simulation results for a cluster of $N = 1944$ spins. These simulations have a native energy resolution of 0.015 meV, set by the length of the time–sequence, $N_t = 400$. Results have further been multiplied by a Gaussian envelope in time, to remove numerical artifacts coming from the finite extent of the time sequence. This is equivalent to a convolution with a Gaussian of FWHM = 0.04 meV in energy, to set a combined limit on the energy resolution of $\Delta\omega = 0.055$ meV. And, in order to compensate for a gradient in intensity $\propto T/\omega$, coming from classical thermodynamics, we do not plot the dynamical structure factor $S(\mathbf{q}, \omega)$ directly, but rather its first moment

$$\tilde{S}(\mathbf{q}, \omega) = \omega S(\mathbf{q}, \omega) \ . \tag{1}$$

Plotted in this way, excitations at intermediate and high energy exhibit similar spectral weight, and it is possible to identify individual bands of dispersing excitations, especially at high values of magnetic field.

It is also interesting to resolve the contributions to $S(\mathbf{q}, \omega)$ coming from transverse and longitudinal fluctuations, i.e.

$$S(\mathbf{q}, \omega) = S^{\perp}(\mathbf{q}, \omega) + S^{\parallel}(\mathbf{q}, \omega) \tag{2}$$

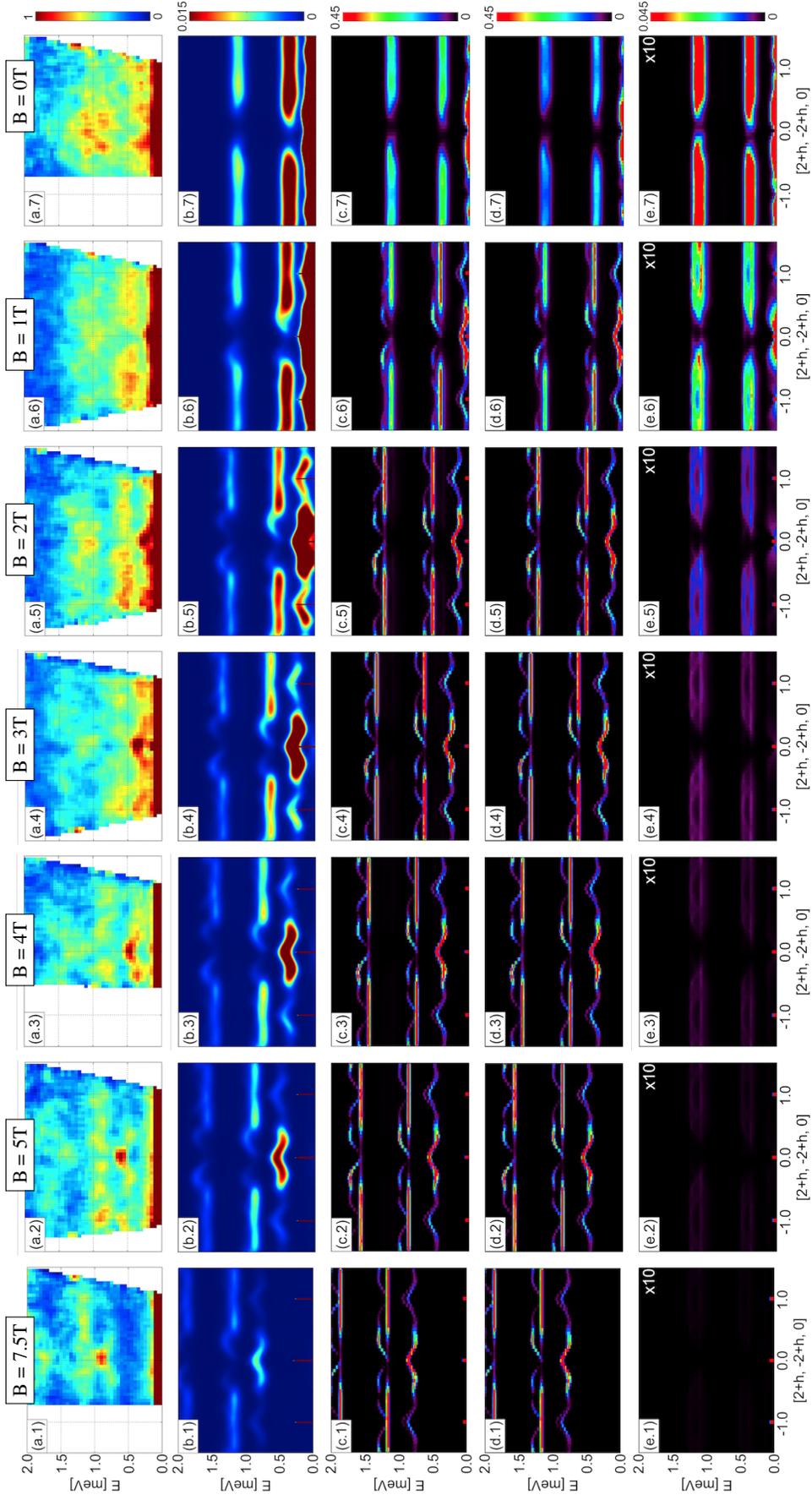

Figure 1. **Evolution of spin excitations as a function of magnetic field.** (a) Inelastic neutron scattering results for $Ca_{10}Cr_7O_{28}$, reproduced from [2], for magnetic fields ranging from $B = 7.5$ T to $B = 0$ T. (b) Predictions for inelastic scattering taken from MD simulations of $\mathcal{H}_{\text{BBK}}$, as described in the Methods section of the main text. (c) Simulation results for the first moment of the dynamical structure factor $\tilde{S}(\mathbf{q}, \omega)$ [Eq. (S1)]. (d) Contribution to $\tilde{S}(\mathbf{q}, \omega)$ coming from transverse spin fluctuations $\tilde{S}^{\perp}(\mathbf{q}, \omega)$ [Eq. (S5)]. (e) Contribution to $\tilde{S}(\mathbf{q}, \omega)$ coming from longitudinal spin fluctuations $\tilde{S}^{\parallel}(\mathbf{q}, \omega)$ [Eq. (S6)], multiplied ×10. Further details of simulations can be found in the text.

where

$$S^\perp(\mathbf{q}, \omega) = \frac{1}{\sqrt{N_t}} \sum_n^{N_t} e^{i\omega\, n\delta t} \langle \mathbf{S}_\mathbf{q}^\perp(t) \cdot \mathbf{S}_{-\mathbf{q}}^\perp(0) \rangle\,, \qquad \mathbf{S}_i^\perp = (S_i^x, S_i^y)\,, \qquad (3)$$

and

$$S^\parallel(\mathbf{q}, \omega) = \frac{1}{\sqrt{N_t}} \sum_n^{N_t} e^{i\omega\, n\delta t} \langle S_\mathbf{q}^z(t) \cdot S_{-\mathbf{q}}^z(0) \rangle\,. \qquad (4)$$

In the fourth row [Fig. S1d], we plot results for the first moment of the dynamical structure factor for transverse excitations

$$\tilde{S}^\perp(\mathbf{q}, \omega) = \omega S^\perp(\mathbf{q}, \omega)\,. \qquad (5)$$

At high values of field, this clearly picks out six different branches spin–wave excitations, in correspondence with the results of linear spin–wave theory [2]. The gap to each of these excitations scales linearly with field, with the gap to the lowest lying transverse excitation closing for $B = 1$ T [cf. Fig. S3]. The intermediate and high–energy spin–wave branches are qualitatively similar, and both contain flat bands of localised spin–wave excitations, similar to those found in the Heisenberg antiferromagnet on the Kagome lattice [3–5].

Meanwhile, in the fifth row [Fig. S1d], we plot results for the first moment of the dynamical structure factor associated with longitudinal excitations

$$\tilde{S}^\parallel(\mathbf{q}, \omega) = \omega S^\parallel(\mathbf{q}, \omega)\,. \qquad (6)$$

The strongest signal in the longitudinal channel is an elastic peak in the zone–center, corresponding to the finite magnetisation of the system for $B > 0$ T. These are visible in Fig. 2h, but are eliminated here by the factor $\omega$ in Eq. (6).

Apart from the elastic signal, the spectral weight found in longitudinal channel becomes negligible for $B > 5$ T. However, at lower values of magnetic field, simulations also reveal three bands of highly–localised longitudinal excitations. The energy of these excitations does not depend on the value of magnetic field, and the lowest–lying longitudinal excitations are found to be gapless. The spectral weight found in all three bands grows as the magnetic field (and thereby the magnetisation) is reduced. The structure of these longitudinal excitations is discussed in more detail below.

### DYNAMICAL STRUCTURE FACTORS AT $B = 2$ T

In Fig. 2 we compare MD results for transverse and longitudinal spin excitations found at a field of $B = 2$ T. Simulations were carried out at $T = 222$ mK, for parameters appropriate to $Ca_{10}Cr_7O_{28}$, as described in the main text. No attempt has been made to correct for the magnetic form factor of $Cr^{5+}$, experimental resolution, or the polarisation–dependence of scattering.

We first consider transverse excitations, as resolved in $S^\perp(\mathbf{q}, \omega)$ [Eq. (S3)]. For this value of field, it is still possible to resolve six distinct bands of spin–wave excitations, grouped into pairs at low, intermediate and high energy [Fig. S2(a)]. These exhibit flat bands at energies of $E = 0.54$ meV and $E = 1.23$ meV. At an energy of $\hbar\omega = 0.57$ meV (slightly above the lowest–lying flat band), results for $S^\perp(\mathbf{q}, \omega)$, reveal characteristic "half–moon" features near the zone centers, overlaid on a lattice of highly–structured "bow–tie" features [Fig. S2(c)]. The origin of these "half–moon" features, which are also observed in MD simulations of the Heisenberg antiferromagnet on a Kagome lattice [6], will be discussed elsewhere [7]. At an energy of $\hbar\omega = 0.54$ meV, the "bow–tie" features resolve into a set of zone–center pinch points, familiar from the Heisenberg antiferromagnet on a Kagome lattice [3–5] [Fig. S2(e)]. Very different structure is resolved in transverse fluctuations at an energy of $\hbar\omega = 0.135$ meV, where the rings, characteristic of a spiral spin liquid are clearly resolved [Fig. S2(g)]. The origin of these features is discussed at some length in the main text.

We now turn to longitudinal excitations, as resolved in simulation results for $S^\parallel(\mathbf{q}, \omega)$ [Eq. S4]. Like transverse spin fluctuations, these are naturally grouped into excitations at low, intermediate and high energy, which show little dispersion [Fig. S2(b)]. Cuts through $S^\parallel(\mathbf{q}, \omega)$ at intermediate energy reveal a network of broad features [Fig. S2(d) and Fig. S2(f)]. Results for high energy excitations (not shown) are very similar. But once again, results at low energy have a dramatically different character, with the majority of spectral weight found in elastic peaks in one quarter of the Brillouin zones, which interpolate through broad, diffuse, scattering to the BZ boundary [Fig. S2(h)].

### FIELD EVOLUTION OF GAP TO TRANSVERSE SPIN EXCITATIONS

In Fig. S3 we show the field–evolution of the gap to the lowest–lying transverse spin excitations, extracted from MD simulation results for $S^\parallel(\mathbf{q}, \omega)$ [cf. Fig. S1]. The gap, $\Delta(B)$, interpolates linearly to zero for $B \to 1.0(5)$ T. This result is consistent with changes in the form of the specific–heat $c(T)$ at $B = 1.0$ T, as reported by Balz *et al.* [2].

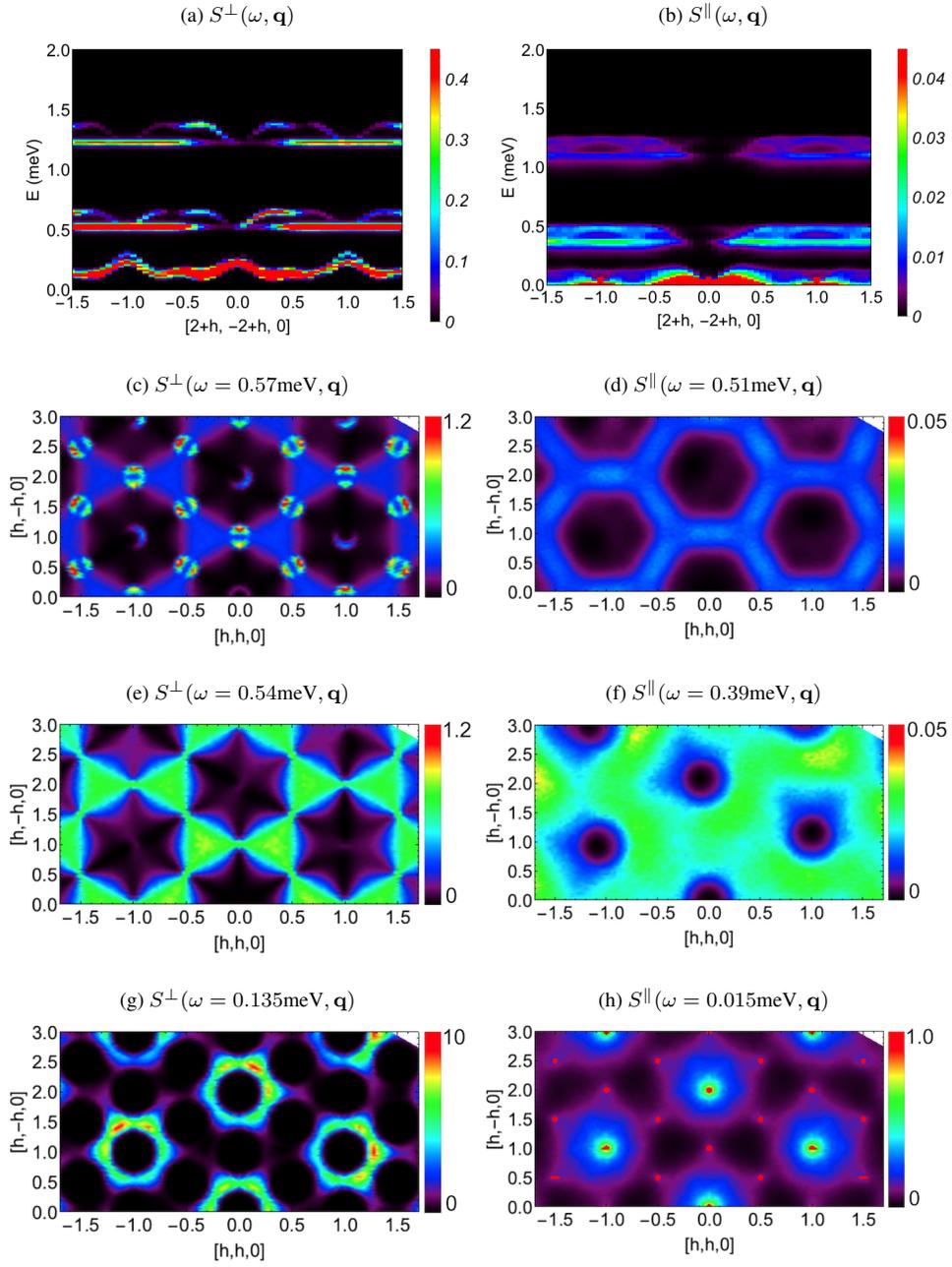

Figure 2. **Transverse and longitudinal spin excitations, as found in MD simulations at $B = 2$ T.** (a) Simulation results for the transverse structure factor $S^\perp(\mathbf{q}, \omega)$ [Eq. (3)], showing six distinct branches of spin–wave excitations. (b) Corresponding results for $S^\parallel(\mathbf{q}, \omega)$ [Eq. (4)], showing weakly–dispersing longitudinal spin excitations at three distinct energy scales. (c) $S^\perp(\mathbf{q}, \omega)$ at intermediate energy $\hbar\omega = 0.57$ meV, showing "half–moon" features overlaid on a lattice of "bow–tie" features. (d) $S^\parallel(\mathbf{q}, \omega)$ at intermediate energy $\hbar\omega = 0.51$ meV, showing a network of broad scattering. (e) $S^\perp(\mathbf{q}, \omega)$ at intermediate energy $\hbar\omega = 0.54$ meV, showing sharp "pinch–point" features. (f) $S^\parallel(\mathbf{q}, \omega)$ at intermediate energy $\hbar\omega = 0.39$ meV, showing evolution of the network of broad scattering. (g) $S^\perp(\mathbf{q}, \omega)$ at low energy $\hbar\omega = 0.135$ meV, showing the ring–feature associated with the spiral spin liquid. (h) $S^\parallel(\mathbf{q}, \omega)$ at low energy $\hbar\omega = 0.015$ meV, showing zone–center elastic peaks associated with finite magnetisation, and accompanying broad, diffuse scattering. All simulations were carried out at $T = 222$ mK, for parameters relevant to $Ca_{10}Cr_7O_{28}$. Further details of simulations can be found in the main text.

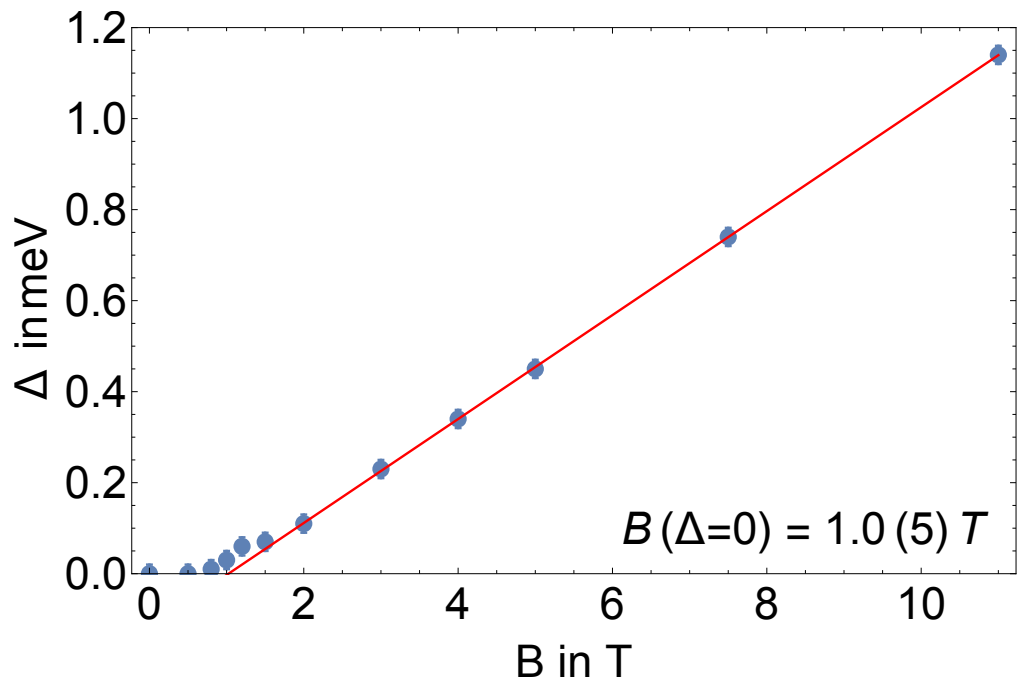

Figure 3. **Field evolution of gap to the lowest–lying transverse spin excitations, $\Delta(B)$**. MD simulations were carried out for parameters relevant to $Ca_{10}Cr_7O_{28}$, at a temperature of $T = 222$ mK, as described in the main text. The red line shows a linear regression of the data, with $\Delta(B) \to 0$ for $B = 1.0(5)$ T. Error bars on points reflect the finite energy resolution of MD results.


\end{document}